\documentclass[aps,prb,twocolumn,superscriptaddress,showpacs]{revtex4}
\usepackage{graphicx}

\begin{document}
\title{Origin of the peaks in the Nernst coefficient of bismuth in
strong magnetic fields}

\author{Yu.~V.~Sharlai}
\affiliation{B.~Verkin Institute for Low Temperature Physics \&
  Engineering, Ukrainian Academy of Sciences,
   Kharkov 61103, Ukraine}

\author{G.~P.~Mikitik}
\affiliation{B.~Verkin Institute for Low Temperature Physics \&
  Engineering, Ukrainian Academy of Sciences,
   Kharkov 61103, Ukraine}

\begin{abstract} We explain the origin of most of the peaks in the Nernst
coefficient that were recently observed at magnetic fields
directed along the trigonal axis and the bisectrix direction in
bismuth. Additional experiments are discussed that enable one to
verify our explanation.
\end{abstract}
\date{\today}

\maketitle

In the recent paper \cite{B1} oscillations of the Nernst
coefficient in bismuth were for the first time observed for the
magnetic fields directed along the trigonal and bisectrix axes of
the crystal. Moreover, several unusual peaks of this coefficient
were discovered for very high magnetic fields $H$ ($14 \le H \le
33$ $T$) parallel to the trigonal axis. \cite{B2} These peaks were
concomitant with quasi-plateaus in the Hall coefficient, and the
authors of Ref.~\onlinecite{B2} suggested that these results are a
signature of an exotic quantum fluid reminiscent of the fluid
associated with the fractional quantum Hall effect. Interestingly,
in the same interval of the high magnetic fields several jumps of
the magnetization were observed which were ascribed to
field-induced instabilities of the ground state of interacting
electrons in bismuth. \cite{Ong}

In this paper we show that positions of the peaks in the Nernst
signal for $H$ along the bisectrix axis and for $H < 12$ $T$
applied along the trigonal axis can be explained using a simple
model of the electron energy spectrum of bismuth in magnetic
fields. In particular, the most of the peaks are due to the hole
Fermi surface of bismuth, while some peaks result from its
electron part. On the basis of our calculation we also predict new
additional peaks that have not yet been observed. However, the
unusual peaks observed in the Nernst signal at high magnetic
fields cannot be explained in this way directly. Nevertheless, we
show that at least some of these peaks can be reproduced if one
assumes that a small deviation of the magnetic-field direction
from the trigonal axis occurred in the experiments. We also
theoretically analyze dependences of the peaks on this tilt angle
of the magnetic field. These angular dependences will enable one
to distinguish between the peaks that appear even in the
one-electron approximation and the peaks that are really due to
new collective effects.

The electron-band structure of bismuth is well known; see, e.g.,
papers \onlinecite{Ed,Pn} and the references cited therein. The
Fermi surface of bismuth consists of one hole ellipsoid located at
the T point of its Brillouin zone, and of three closed electron
surfaces of nearly ellipsoidal shape centered at the points L. The
spectrum of the holes in bismuth is well approximated by the
simple parabolic model.\cite{Ed} On the other hand, the electron
spectrum near the point L is accurately described by the model of
McClure, \cite{Mc} and the parameters of this model are well
known. \cite{Pn,Bi} However, in the framework of the McClure model
the spectrum of the electrons {\it in a magnetic field} $H$ can be
found analytically only if $H$ is directed along the longest axis
of the electron ellipsoid, i.e., if ${\bf H}$ practically
coincides with the bisectrix axis. \cite{Bi} To find the Landau
levels of the electrons for an arbitrary direction of $H$, two
empirical expressions for these levels were proposed many years
ago. \cite{Sm,Vec} These expressions permitted one to describe a
number of experimental data. \cite{Sm,Vec,HirumaMiura,Bompadre}
Below we shall use the expression suggested in
Ref.~\onlinecite{Sm} to analyze the oscillations of the Nernst
coefficient.

The electron ellipsoid at the point L is elongated along the
bisectrix axis. If one describes this ellipsoid by the so-called
model of Lax \cite{lax} in which only linear terms in
quasimomentum $p$ are taken into account in the electron
Hamiltonian, the electron spectrum in the magnetic field $H$ can
be found analytically at any direction of $H$. But the Lax model
cannot accurately describe the electron ellipsoid along its
elongation, and that is why McClure \cite{Mc} took into account
also quadratic terms in $p$ for this direction in his Hamiltonian.
It is these terms that do not permit one to find the spectrum
analytically at the arbitrary directed magnetic fields within the
McClure model.

\begin{figure}
\includegraphics[scale=0.9]{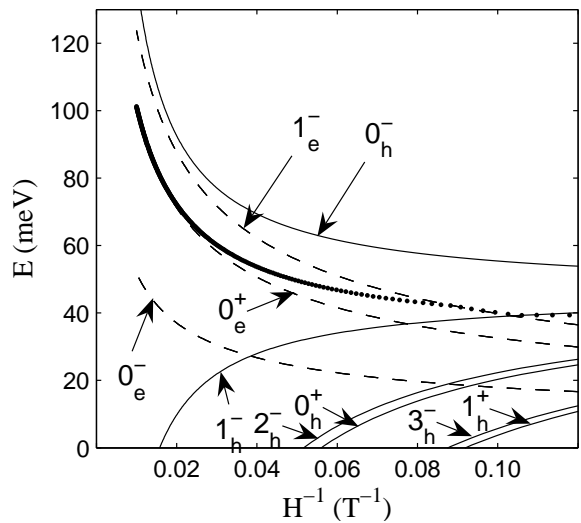}
\caption{\label{fig1} The $H$-dependence of the Landau levels for
the electrons (the dashed lines) at the point L and for the holes
(the solid lines) at the point T in bismuth. The dependence of the
Fermi level on $H$ is also shown (the line with dots). The
symbols near the lines indicate the Landau-level numbers and the
direction of the spin projection on the magnetic field. The
subscripts e and h stand for the electrons and holes.
 } \end{figure}   

To allow for the deviation of the real electron spectrum from the
Lax model, Smith, Baraff and Rowell \cite{Sm} suggested a simple
generalization of the formula that describes the Landau levels in
the model of Lax. According to Ref.~\onlinecite{Sm}, in the
presence of the magnetic field $H$ the $n$th Landau level $E_n$
for an electron with the quasimomentum $p_H$ along ${\bf H}$ can
be found from the equation:
\begin{equation}\label{1}
E\left (1+{E\over E_G}\right ) = (n+{1\over 2})\hbar \omega_c
+{p_H^2\over 2m_H}\pm {1\over 2}g\beta_0H,
\end{equation}
where signs $\pm$ correspond to the electron spins that are
antiparallel and parallel to ${\bf H}$, respectively; the energy
$E$ is measured from the edge of the conduction band; $\omega_c$
is the cyclotron frequency
\[
\omega_c={eH\over m_c c},
\]
$E_G$ is the gap between the conduction and valence bands at the
point L; $g$ is the effective electron g factor at this point;
$\beta_0$ is the Bohr magneton, and the longitudinal and the
cyclotron masses, $m_H$ and $m_c$, are given by
\begin{eqnarray}\label{2}
 m_H&=& {\bf h}\cdot {\bf m}^e\cdot {\bf h}, \\
 m_c&=&[\det{\bf m}^e/m_H]^{1/2}. \label{3}
\end{eqnarray}
Here ${\bf h}$ is the unit vector in the direction of the magnetic
field $H$. The effective mass tensor ${\bf m}^e$ has the form:
\begin{equation}\label{4}
{\bf m}^e= \left (
    \begin{array}{ccc}
    m_{11} & 0 & 0 \\
    0 & m_{22} & m_{23} \\
    0 & m_{23} & m_{33}
    \end{array}
\right ),
\end{equation}
where the axes 1 and 3 coincide with the binary and the trigonal
axes, respectively, while the axis 2 is along the bisectrix
direction. The effective $g$ factor,
\begin{equation}\label{5}
g^2= 4m^2{{\bf h}\cdot {\bf m}^e_s\cdot {\bf h}\over \det{\bf
m}^e_s },
\end{equation}
is defined in terms of a spin-mass tensor ${\bf m}^e_s$ that has
the form similar to Eq.~(\ref{4}). Within the Lax model it is
obtained that the spectrum is described by formulas (\ref{1}) -
(\ref{5}), and ${\bf m}^e_s$ exactly coincides with ${\bf m}^e$,
while Smith, Baraff, and Rowell \cite{Sm} admitted that the
elements of ${\bf m}^e_s$ may differ from the elements of ${\bf
m}^e$ and that they are free parameters. This is just the
generalization proposed in Ref.~\onlinecite{Sm}.

\begin{table}
\caption{\label{table1} Parameters of the Smith-Baraff-Rowell
spectrum\cite{Sm}}
\begin{ruledtabular}
\begin{tabular}{lcccc}
Electrons&$m_{11}$&$m_{22}$&$m_{33}$&$m_{23}$\\ \hline Orbital
mass & 0.00113 & 0.26 & 0.00443 & -0.0195\\ Spin mass & 0.00101 &
2.12 & 0.0109 & -0.13\\ \hline Holes&&$M_1=M_2$&&$M_3$\\ \hline
Orbital mass&&0.07\cite{ex1}&&0.69\\ Spin mass&&0.033&&200\\
\hline $E_g=15.3\,{\rm meV}$ &&& $E_0=38.5\,{\rm meV}$
\end{tabular}
\end{ruledtabular}
\end{table}

Since the spectrum of the holes at the point T is parabolic, the
Landau levels for them can be easily found \cite{Ed}
\begin{equation}\label{6}
E_0- E = (n+{1\over 2})\hbar \omega_c +{p_H^2\over 2m_H}\pm
{1\over 2}g\beta_0H,
\end{equation}
where $E_0$ is the edge of the hole band at this point of the
Brilloun zone. The cyclotron frequency $\omega_c$, the masses
$m_c$ and $m_H$, and the $g$ factor are defined by the same
formulas (\ref{2}), (\ref{3}), (\ref{5}) as for the electrons, but
now the tensor of the effective masses for the holes ${\bf m}^h$
has the form:
\begin{equation}\label{7}
{\bf m}^h= \left (
    \begin{array}{ccc}
    M_1 & 0 & 0 \\
    0 & M_2 & 0 \\
    0 & 0 & M_3
    \end{array}
\right ),
\end{equation}
and a similar expression is valid for the spin-mass tensor, ${\bf
m}_s^h$. All the parameters in Eqs.~(\ref{1}) - (\ref{7}) are
known for bismuth; \cite{Sm} see Table \ref{table1}.

In our subsequent discussion we shall denote the Landau levels
$E_n(p_H=0)$ as $n_{e,h}^{\mp}$ where $n$ is the quantum number of
the level, the subscripts ${e,h}$ stand for electron and holes,
and the signs $\mp$ correspond to the electron spin directed up
and down, respectively. Figure \ref{fig1} shows the $H$-dependence
of these levels for the electrons and holes in bismuth in the case
of the magnetic field directed along the trigonal axis. In this
figure we also show the $H$-dependence of the Fermi level $\mu$
that is found from the equality of the concentrations of the
electrons and holes in bismuth. Since the electron levels that are
below $\mu(H)$ and the hole levels lying above $\mu(H)$ are
filled, one can easily trace the change in the population of the
levels with increasing $H$.

\begin{figure}
\includegraphics[scale=1]{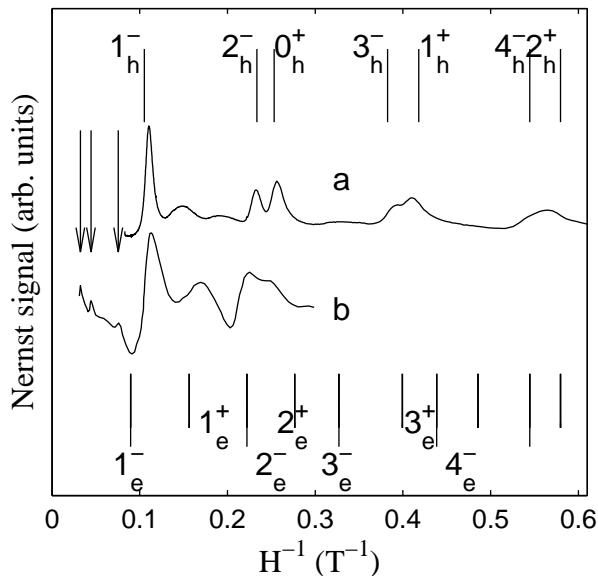}
\caption{\label{fig2} The Nernst signal in the case of the
magnetic field directed along the trigonal axis. The curve {\bf a}
reproduces the experimental data of Ref.~\onlinecite{B1}, while
the curve {\bf b} presents the data of Ref.~\onlinecite{B2} obtained
with the same crystal. The vertical lines mark the calculated
magnetic fields at which the appropriate electron and hole Landau
levels cross the Fermi energy. The notation of the Landau levels
is the same as in Fig.~\ref{fig1}. The arrows show positions of
the unusual peaks \cite{B2} seen in the curve {\bf b}.
} \end{figure}

Figure \ref{fig2} shows the magnetic fields at which the
calculated electron and hole Landau levels cross the Fermi energy.
This figure also shows the experimental Nernst signal\cite{B1,B2}
for low temperatures $T \sim 1$ K. It is seen  that the
experimental peaks with large amplitudes are caused by the hole
ellipsoid, and their positions are well reproduced by our
calculation. At the magnetic fields $H < 2.5$ T the electron
ellipsoids practically do not manifest themselves in the
oscillations of the Nernst coefficient due to the low mobility of
the electrons as compared to the mobility of the holes.\cite{B1}
Only relatively small maxima seen in the interval $2.5$ T $<H< 10$
T can be attributed to the electrons. But the calculated positions
of the electron peaks do not agree accurately with the
experimental data. Moreover, the positions of the small
experimental maxima varies from sample to sample and from
experiment to experiment. \cite{Bpr} However, the most essential
point is that the calculation does not reveal any peaks at the
magnetic fields higher than 11 T. When $H>11$ T, the only hole
level $0_h^{-}$ and the two electron levels $0_e^{+}$ and
$0_e^{-}$ are filled, and these levels do not cross $\mu(H)$ with
increasing $H$.

Of course, one should keep in mind that the above formulas for the
electron Landau levels are not exact. In the case of high magnetic
fields corrections to these formulas were studied in
Refs.~\onlinecite{Vec,HirumaMiura}. However, our analysis shows
that such corrections can only partly improve the situation,
e.~g., it is possible to fit the electron energy levels $1_e^{+}$
and $2_e^{-}$ so that to describe accurately electron maxima
between the hole peaks $1_h^{-}$ and $2_h^{-}$ in the curve a, 
but these corrections cannot describe the unusual peaks 
observed in the magnetic fields above $11$ T.

To resolve this problem, we assume that in the experiments
\cite{B1,B2} the magnetic field was slightly tilted away from the
trigonal axis. Then, the following explanation of these
experimental data is possible: There are three electron ellipsoids
in bismuth, and their energy levels coincide only if the magnetic
field is directed strictly along the trigonal axis. If the
magnetic field begins to tilt away from this axis, each electron
Landau level in Fig.~\ref{fig1} splits into two or three levels
(this depends on the plane of the tilt). The electron energy level
$0_e^{+}$ closely approaches $\mu(H)$ in high magnetic fields,
Fig.~\ref{fig1}. When the tilt splits this level, the levels
resulting from $0_e^{+}$ can intersect $\mu(H)$ in the magnetic
fields $H>11$ T.

\begin{figure}
\includegraphics[scale=1]{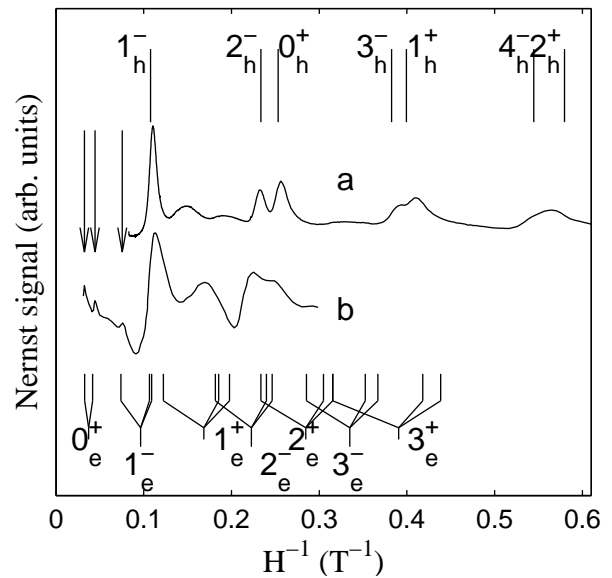}
\caption{\label{f3} The same as Fig.~\ref{fig2}, but with the
magnetic field slightly tilted away from the trigonal axis. Here
the magnetic-field direction ${\bf h}=(\sin\theta\cos\psi,
\sin\theta\sin\psi,\cos\theta)$ is given by $\theta=2.5^\circ$,
$\psi=85^\circ$. We also show the splitting of the electron Landau
levels.
} \end{figure}   

\begin{figure}
\includegraphics[scale=0.9]{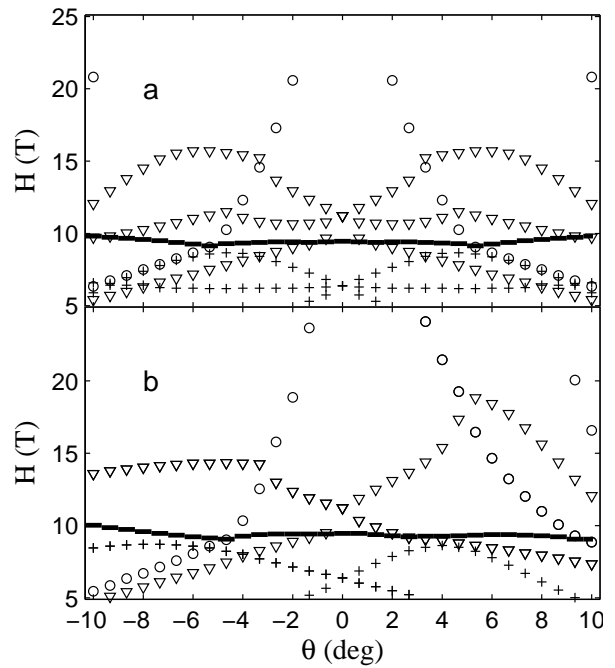}
\caption{\label{f4} The calculated angular dependences of the
magnetic fields, $H(\theta)$, at which the electron and hole
Landau levels cross the Fermi energy; $\theta$ is the angle
between the magnetic field and the trigonal axis. The magnetic
field tilts either towards the binary axis (a) or towards the
bisectrix axis (b). The positions of the hole peak $1_h^{-}$ are
shown by the thick lines, while positions of the electrons peaks
$0_e^{+}$, $1_e^{-}$, and $1_e^{+}$ are marked by the circles,
triangles, and crosses, respectively.
} \end{figure}   

\begin{figure}
\includegraphics[scale=0.9]{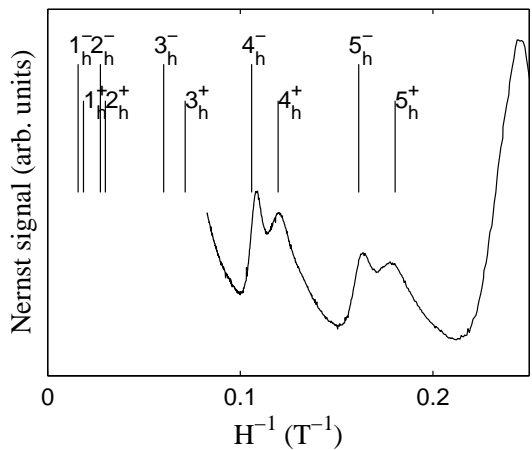}
\caption{\label{f5} The Nernst signal in the case of the
magnetic field directed practically along the bisectrix
axis.\cite{B1} The vertical lines show the calculated magnetic
fields at which the appropriate Landau levels of the holes cross
the Fermi energy. The magnetic-field direction ${\bf
h}=(\sin\theta\cos\phi, \sin\theta\sin\phi, \cos\theta)$ with
$\theta=92^\circ$, $\psi=85^\circ$ is chosen so that the
Landau-level spacing leads to the observed period \cite{B1} of the
oscillations.
} \end{figure}   

To be specific, consider the case when the magnetic-field
direction ${\bf h}=(\sin\theta\cos\psi,
\sin\theta\sin\psi,\cos\theta)$ is given by the angles
$\theta=2.5^\circ$, $\psi=85^\circ$. In this case each electron
level splits into three levels, and the calculated intersections
of these splitted levels with $\mu(H)$ are shown in Fig.~\ref{f3}.
Now the two levels resulting from $0_e^{+}$ indeed intersect
$\mu(H)$ at the magnetic fields that are close to the positions of
the two unusual peaks seen in the curve b [the third level does
not cross $\mu(H)$]. On the other hand, the intersection of one of
the levels $1_e^{-}$ practically coincides with the position of
third unusual peak of this curve. Thus, it is quite possible that
at least some of the unusual peaks can result from the
intersection of the chemical potential with the Landau levels
splitted by a tilt of the magnetic field. A small uncontrollable
tilt of the magnetic field can also explain why the positions of
the electron peaks varies from experiment to experiment.\cite{Bpr}
To verify this explanation and to find out which of the unusual
peaks are really due to some nontrivial physics, it would be
useful to measure angular dependences of their positions. In
Fig.~\ref{f4} we show intersections of some Landau levels with
$\mu(H)$ in the $\theta$-$H$ plane. Interestingly, the upper part
of our figure \ref{f4} is closely reminiscent of Fig.~3 in
Ref.~\onlinecite{Ong}. In particular, the angular dependences of
the magnetic fields at which the Landau-levels $0^{+}_e$ and
$1^{-}_e$ cross the chemical potential $\mu(H)$ are close to the
transition lines $H_2(\theta)$ and $H_1(\theta)$ experimentally
defined in Ref.~\onlinecite{Ong}.

We also calculate the Landau levels when the magnetic field is
close to the bisectrix direction. The obtained results are presented
in Fig.~\ref{f5}. For this geometry in the magnetic fields $H >
6$ T all the electrons are in the lowest Landau level, and the
oscillations of the Nernst signal \cite{B1} shown in
Fig.~\ref{f5} are due to the holes. Note that in the region of
high magnetic fields the calculation predicts additional peaks in
the Nernst signal which have not been observed yet.

In conclusion, the Smith-Baraff-Rowell model \cite{Sm} is
sufficient to explain the origin of most of the peaks in the
Nernst coefficient that were recently observed in bismuth.
\cite{B1,B2} It seems that the positions of some unusual peaks
discovered at very high magnetic fields oriented along the
trigonal axis can be understood under assumption that the magnetic
field was slightly tilted away from the trigonal axis in these
experiments. This assumption can be verified by analyzing
positions of these peaks in tilted magnetic fields. These angular
dependences will enable one to select the peaks that are really
due to a nontrivial physics. We calculate these angular
dependences for the magnetic fields ${\bf H}$ lying in the two
principle planes containing the trigonal axis. The results
obtained in one of the planes are reminiscent of the recent
experimental data. \cite{Ong} Our calculation also predicts
additional high-field peaks in the Nernst signal at the magnetic
field oriented along the bisectrix direction.

We thank K.~Behnia for useful discussions and for providing us
with the experimental data of Fig.~2 and 3. 
This work was supported 
by the France-Ukraine program of scientific collaboration (EGIDE, PHC DNIPRO).


\begin{thebibliography}{}
\bibitem{ex1} This value differs from $0.064$ presented in
Ref.~\onlinecite{Sm}, but it is taken to fit the period of the
oscillations ($0.147 T^{-1}$) observed in
Refs.~\onlinecite{B1,Bompadre}.
\end{thebibliography}
\end{document}